\documentclass[twoside,12pt]{article}
\usepackage{epsfig}
\usepackage[utf8]{inputenc}
\usepackage{amsmath}

\newcommand{\be}{\begin{equation}}
\newcommand{\ee}{\end{equation}}
\newcommand{\bea}{\begin{eqnarray}}
\newcommand{\eea}{\end{eqnarray}}

\begin{document}

\title{Scalar susceptibilities and Electromagnetic thermal mass differences in Chiral Perturbation Theory }
\author{R. Torres Andrés,\, A. Gómez Nicola\\ Departamento de Física Teórica II, Universidad Complutense de Madrid, Spain}
\date{}
\maketitle
\begin{abstract}

We make a thermal analysis of the light scalar susceptibilities using SU(3)-Chiral Perturbation Theory to one loop taking into account the QCD source of isospin breaking (IB), i.e corrections coming from $m_u\neq m_d$. The value of the connected scalar susceptibility in the infrared regime, the one relevant when approaching chiral symmetry restoration, and below the critical temperature is found to be entirely dominated by the $\pi^0-\eta$ mixing, which leads to model-independent $\mathcal{O}(\epsilon^0)$ corrections, where $\epsilon\sim m_d-m_u$, in the combination $\chi_{uu}-\chi_{ud}$ of flavour breaking susceptibilities.
We also present preliminary results for the corrections to the real part of the pion self-energy at nexst-to-leading order in SU(2)-Chiral Perturbation Theory taking into account electromagnetic interaction. The zero and finite temperature results for the charged and neutral pions are given in terms of the three-momentum of the external pion; and their difference is calculated to this order stressing the fact that, at low and moderate temperature, the mass splitting $M_{\pi^\pm}-M_{\pi^0}$ grows with temperature for, at least, non-zero charged pion mass running inside the loops. 
\end{abstract}

\section{Introduction}

The low-energy sector of QCD has been successfully described over recent years within the chiral lagrangian framework. Chiral Perturbation Theory (ChPT) is based on the spontaneously breaking of chiral symmetry $SU_V(N_f)\times SU_A(N_f)\rightarrow SU_V(N_f)$ with $N_f=2,3$ light flavors and provides a consistent, systematic and model-independent scheme to calculate low-energy observables \cite{we79,Gasser:1983yg,Gasser:1984gg}.  The effective ChPT lagrangian is constructed as an expansion of the form ${\cal L}={\cal L}_{p^2}+{\cal L}_{p^4}+\dots$ where $p$ denotes a meson energy scale compared to the chiral scale $\Lambda_{\chi}\sim$ 1 GeV. 
The formalism can also be extended to  finite temperature $T$, in order to describe  meson gases and their evolution towards chiral symmetry restoration for $T$ below the critical temperature $T_c$ \cite{Gasser:1986vb,Gerber:1988tt}, where $T_c\simeq$ 180-200 MeV from lattice simulations \cite{Bernard:2004je,Aoki:2009sc,Cheng:2009zi}. The use of ChPT in this context is important in order to provide model-independent results for the evolution of the different observables with $T$, supporting the original predictions for chiral restoration \cite{PisWil84}, also confirmed by lattice simulations, which are consistent with a crossover-like transition for $N_f=3$ (2+1 flavours in the physical case), a second-order one for $N_f=2$ in the $O(4)$ universality class and first order in the degenerate case of three equal flavours.

   The invariance under $SU_V(2)$ vector group is the isospin symmetry, which is a very good approximation to Nature. However, there
 are several proccesses where isospin breaking corrections are phenomenologically relevant, for example sum rules for quark condensates \cite{Gasser:1984gg}, meson masses  \cite{Urech:1994hd} or  pion scattering \cite{Knecht:1997jw,Meissner:1997fa}.
 There are two possible sources of isospin breaking: the QCD $m_d-m_u$ light quark mass difference and electromagnetic interactions. Both can be accommodated within the ChPT framework. From the first source we expect corrections of order $(m_d-m_u)/m_s$, encoded in the quark mass matrix,  which generates also a  $\pi^0\eta$ mixing term in the $SU(3)$ lagrangian \cite{Gasser:1984gg}.  On the other hand, the  electromagnetic interactions are included in the ChPT effective lagrangian via the external source method and give rise to new terms \cite{Urech:1994hd,Knecht:1997jw,Meissner:1997fa,Ecker:1988te} of order ${\cal L}_{e^2}$, ${\cal L}_{e^2p^2}$ and so on, being $e$ the electric charge. It is possible to accomodate these terms into the ChPT power counting scheme by considering formally $e^2=\mathcal{O}(p^2/F^2)$, with $F$ the pion decay constant in the chiral limit.

The purpose of this paper is to calculate the leading thermal contributions to the connected and disconnected scalar susceptibilities taking into account isospin breaking, and to show our preliminary results about the thermal evolution of the masses of the charged and neutral pions.

\section{Light scalar susceptibilities and the role of the $\pi^0-\eta$ mixing}

Light quark condensates, $\langle\bar uu\rangle$ and $\langle\bar dd\rangle$, at zero and finite temperature  have been calculated to one loop in  \cite{Nicola:2010xt} and \cite{Nicola:2011gq}, respectively, within the framework of SU(3)-ChPT taking into account both sources of IB. The main feature we want to stress is that there is a $\pi^0-\eta$ mixing term appearing through the tree-level mixing angle $\varepsilon$ defined by $\tan 2\varepsilon=\frac{\sqrt{3}}{2}\frac{m_d-m_u}{m_s-\hat m}$.

Different light quark masses allow to consider three independent light scalar susceptibilities defined as 
\begin{equation}
\chi_{ij}=-\frac{\partial}{\partial m_i}\langle\bar q_j q_j\rangle=\frac{\partial^2}{\partial m_i\partial m_j} \log Z(m_u\neq m_d).
\label{indepsus}   
\end{equation}

From now on in this section, we will neglect the electromagnetic corrections because they are small and not relevant for our present discussion, so we will put $e=0$. Then, to leading order in the mixing angle, the  contribution of the $\pi^0-\eta$ mixing in the quark condensate sum is of order $\epsilon^2$ whereas for the difference it goes like $\varepsilon$. The thermal functions $g_i(T,M_i),\, i=\pi^0, \eta;$ defined as $\!g_i(T)\!=\frac{1}{4\pi^2F^2}\int_0^\infty dp \frac{p^2}{E_p} \frac{1}{e^{\beta E_p}-1},$  with $E_p^2=p^2+M_i^2$ and $\beta=T^{-1}$; are suppressed by those coefficients and the quark condensates do not receive important corrections. The important point is that differentiating with respect to a light quark mass is essentially the same as differentiating with respect to $\varepsilon\sim \frac{m_d-m_u}{m_s}$,
so the suppression of the thermal functions is smaller in the case of the susceptibilities than in the quark condensate. 

Because of the linearity in $\varepsilon$ in $\langle uu-dd\rangle$ for a small mixing angle, the combinations $\chi_{uu}-\chi_{ud}$ and $\chi_{dd}-\chi_{du}$ receive an $\mathcal{O}(1)$ IB correction due to $\pi^0-\eta$ mixing, which would not be found if $m_u=m_d$ is taken from the beginning. The analysis of the $\varepsilon$-dependence  of $\langle uu-dd\rangle$ shows that, up to $\mathcal{O}(\epsilon)$, $\chi_{uu}\simeq \chi_{dd}$, so combinations like $\chi_{uu}-\chi_{dd}$, which also vanish with $m_u=m_d$, are less sensitive to IB.

One can also relate these flavour breaking susceptibilities with the connected and disconnected ones \cite{Smilga:1995qf}, often used in lattice analysis \cite{Ejiri:2009ac,DeTar:2008qi}: $\chi_{dis}=\chi_{ud}$, and $\chi_{con}=\frac{1}{2}(\chi_{uu}+\chi_{dd}-2\chi_{ud})$.
From the previous analysis, we get $\chi_{con}\simeq\chi_{uu}-\chi_{ud}$. 

Therefore, our model-independent analysis including IB effects provides the leading nonzero contribution for the connected susceptibility which arises partially from $\pi^0-\eta$ mixing. This is particularly interesting for the lattice, where artifacts such as taste-breaking mask the behaviour of $\chi_{con}$ with the quark mass and $T$ when approaching the continuum limit \cite{DeTar:2008qi}. 
In fact, our ChPT approach is useful to explore the chiral limit ($m_{u,d}\rightarrow 0$) or infrared (IR) regime, which gives a qualitative picture of the behaviour near chiral symmetry restoration. In this regime $M_\pi\ll T\ll M_K$, and therefore we can neglect thermal heavy particles, which are exponentially suppresed.

The leading order results for the connected and disconnected susceptibilities at zero temperature are the following

\begin{equation}\label{con}
\chi^{IR}_{\text{con}}(T=0)= 8B_0^2[2L_8^r(\mu)+H_2^r(\mu)])-\frac{B_0^2}{16\pi^2}\left(1+\log\frac{M^2_K}{\mu^2}\right)-\frac{B_0^2}{24\pi^2}\log\frac{M^2_\eta}{\mu^2}+\mathcal{O}(\epsilon^2)
\end{equation}

\begin{equation}\label{dis}
\chi^{IR}_{\text{dis}}(T=0)=32B_0^2L_6^r(\mu)-\frac{3B_0^2}{32\pi^2}\left(1+\log{\frac{M^2_{\pi}}{\mu^2}}\right)+\frac{B^2_0}{288\pi^2}\left(5\log\frac{M^2_\eta}{\mu^2}-1\right)+\mathcal{O}(\epsilon^2),
\end{equation}

where $B_0$ is the parameter which relates masses and quark condensates at tree level via the Gell Mann-Oakes-Renner formula, and $L_6$, $L_8$, $H_2$ are low energy constants.

The log term in equation (\ref{dis}) is the dominant at $T=0$ and can be found in \cite{Smilga:1995qf}, but the connected IR susceptibility (\ref{con}) is not zero at $T=0$, because it receives contributions of order $\mathcal{O}(1)$ in the mixing angle.

If we consider the pion gas in a thermal bath, then expressions (\ref{con})-(\ref{dis}) are modified according to
\begin{equation}
\label{conT}
\left[\chi_{\text{con}}(T)-\chi_{\text{con}}(0)\right]^{IR}=\frac{B_0^2}{18}\frac{T^2}{M_\eta^2}+\mathcal{O}\left(\epsilon^2\, B^4_0\,\frac{T^4}{M^4_\eta}\right)+\mathcal{O}\left(\exp\left[-\frac{M_{\eta, K}}{T}\right]\right),
\end{equation}

\begin{equation}
\label{disT}
\left[\chi_{\text{dis}}(T)-\chi_{\text{dis}}(0)\right]^{IR}=\frac{3B^2_0}{16\pi}\frac{T}{M_\pi}+\mathcal{O}\left(\epsilon^2\, B^4_0\,\frac{T^4}{M^4_\eta}\right)+\mathcal{O}\left(\exp\left[-\frac{M_{\eta, K}}{T}\right]\right).
\end{equation}

Note that, as we have already mentioned, the eta mass term in equation (\ref{conT}) and in the subleading corrections in the mixing angle comes from the $\epsilon$-analysis and the IR expansion of the $g_1(M_\pi)$, and does not have anything to do with thermal etas.

\begin{figure}
\includegraphics[scale=0.75]{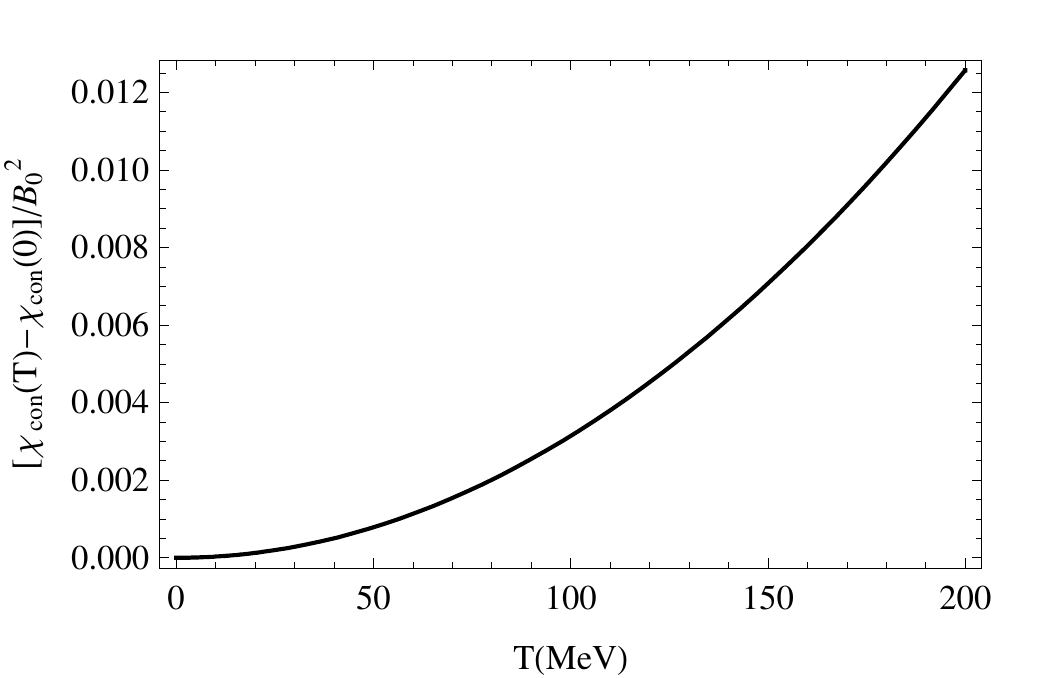}
\centering
\caption{Connected IR susceptibility normalized to $B^2_0$, for fixed tree level eta mass and $m_s=80$ MeV.}
\label{fig:figa}
\end{figure}

\begin{figure}
\includegraphics[scale=0.75]{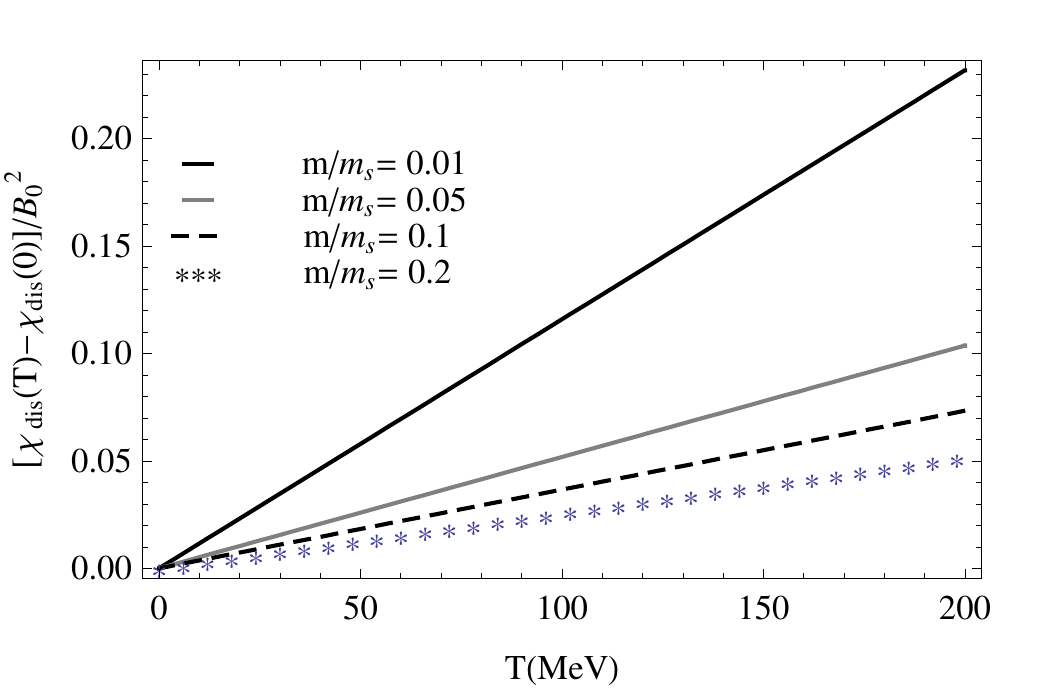}
\centering
\caption{Disconnected IR susceptibility normalized to $B^2_0$, for several light quark mass ratios and fixed tree level eta mass ($m/m_s=0,05$ is the physical case).}
\label{fig:figb}
\end{figure}

Figure \ref{fig:figa} and \ref{fig:figb} show, respectively, the connected susceptibility (\ref{conT}) for fixed tree level eta mass (proportional to $\sqrt{B_0\, m_s}$ in the IR regime), and the disconnected one (\ref{disT}) for several values of the light quark mass ratio $m/m_s$, and also with fixed tree level eta mass.
The leading scaling with $T$ and the light quark mass in this regime for the disconnected piece goes like $\frac{T}{\sqrt{m}}$, i.e the same scaling calculated in \cite{Smilga:1995qf,Ejiri:2009ac}; whereas the connected susceptibility grows quadratic in $T$ over a mass scale much greater than the SU(2) Goldstone boson's one. Therefore, in the continuum limit, we only expect $\chi_{dis}$ to peak near the transition, as the $m\rightarrow 0^+$ limit in figure \ref{fig:figb} clearly shows.

\vspace*{4mm}

\section{Thermal charged and neutral pion masses in SU(2)-ChPT at $\mathcal{O}(p^4)$}

\begin{figure}[h!t]
\centering
\includegraphics[scale=0.14]{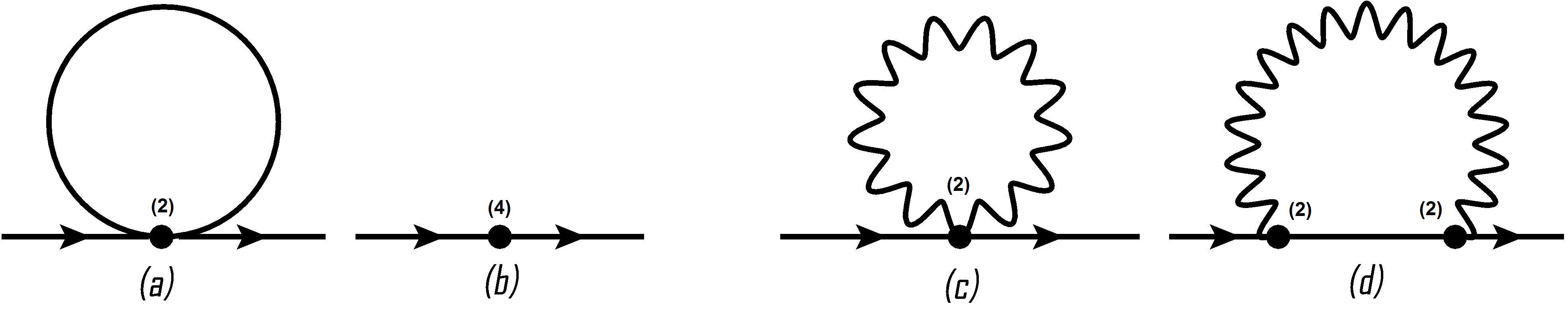} 
\label{fig:diagrams}
\caption{Diagrams for the pion self-energy at order $\mathcal{O}(p^4)$. (a), (b) represent pion tadpole contributions and the tree level NLO diagram wich renormalizes the loops, respectively. (c) (photon tadpole) and (d) (one-photon exchange) only correct the charged pion self-energy.}
\end{figure}

If we consider virtual photons in the calculation of the real part of the self-energy in the mass shell, there are four relevant diagrams that correct the masses at order $\mathcal{O}(p^4)$: there are pion tadpoles, diagram(a) in Figure \ref{fig:diagrams}, and the tree level NLO diagram needed for renormalization, (b), where both charged and neutral pions participate; and diagrams with virtual photons, (c) and (d), which only modify the charged pion mass. 

The photon-tadpole diagram (c) is proportional to the photon mass and therefore vanishes at zero temperature, while pion tadpoles (a) and the photon-exchange diagram (d) are finite and chiral scale-independent once regularized and combined with diagram (b). 
The LO corrections to the masses of the SU(2) NGB are calculated as $M^2=\hat M^2+\Sigma(\hat M^2),$ where $\hat M$ is the respective tree level mass.  At zero temperature the neutral and charged pion masses are \cite{Knecht:1997jw}
 \begin{equation}
M^2_{\pi^0}=\hat M^2_{\pi^0}\left(1+2\mu_{\pm}-\mu_{0}+e^2\mathcal{K}_{\pi^0}+2l_3^r\frac{\hat M^2_{\pi^0}}{F^2}\right)-2\frac{B^2}{F^2}l_7^r(m_d-m_u)^2-\frac{4}{3}Be^2 k_7(m_d-m_u)
\end{equation}
where $\mu_{\pm,0}=\frac{M^2_{\pi^\pm,\pi^0}}{32\pi^2F^2}\log\frac{M^2_{\pi^\pm,\pi^0}}{\mu_\chi^2},$ and $\mathcal{K}_{\pi^0}=-\frac{20}{9}\left[k_1^r+k_2^r-\frac{9}{10}(2k_3^r-k_4^r)-k_5^r-k_6^r-\frac{1}{5}k_7^r\right],$ being $k_i$  electromagnetic low energy constants, $Z$ the parameter that corrects the leading order charged pion mass; and
 \begin{equation}
 M^2_{\pi^\pm}\!\!=\!\hat M^2_{\pi^0}\left[1+\frac{e^2}{4\pi}+\mu_{0}+e^2\mathcal{K}_{\pi^\pm}^A+2l_3^r\frac{\hat M^2_{\pi^0}}{F^2}\right]\!+\!2e^2F^2\left[Z\left(1+\frac{e^2}{4\pi}\right)\!+\!e^2\mathcal{K}_{\pi^\pm}^B\!-\!(3+4Z)\mu_{\pm}\right]\!-\!\frac{4}{3}Be^2 k_7(m_d-m_u), \end{equation}
with the definitions $$\mathcal{K}^A_{\pi^\pm}=-\frac{20}{9}\left[k_1^r+k_2^r-k_5-\frac{1}{5}(23k_6^r+18k_8^r+k_7^r)\right],$$
and $$\mathcal{K}^B_{\pi^\pm}=-\frac{10}{9}\left[2Z(k_1^r+k_2^r)-\frac{1}{2}k_{13}-k_{14}\right].$$

There is a factor $1/2$ in the coefficient of the $k_7$ for both masses wich does not appear in \cite{Knecht:1997jw}, and that was also noted by \cite{Schweizer:2002ft}.

To this point the total mass difference between charged and neutral pions becomes
\begin{eqnarray}
M^2_{\pi^\pm}-M^2_{\pi^0}=2\hat M^2_{\pi^0}(\mu_0-\mu_\pm)+2\frac{B^2}{F^2}l_7^r(m_d-m_u)^2+2e^2F^2\left[Z\left(1+\frac{e^2}{4\pi}\right)\right.\nonumber\\ \left.+e^2\mathcal{K}^B_\pm\right]+\hat M^2_{\pi^0} e^2\left[\frac{1}{4\pi}+\mathcal{K}_\pm^A-\mathcal{K}_{\pi^0}\right]-2e^2F^2(3+4Z)\mu_\pm,
\end{eqnarray}
which has pure strong, pure EM and mixed EM-strong contributions.

If the pions are immersed in a thermal bath, there appear new contributions with no UV divergences, since these only appear in the zero temperature part and they have been already renormalized. The pion tadpoles, charged or not, give rise to $g_1(M,T)$ thermal functions. For the neutral pion mass we get

\begin{equation}
M^2_{\pi^0}=\hat M^2_{\pi^0}(T=0)\left[1+\frac{1}{F^2}\left(g_1(M^2_{\pi^\pm},T)-\frac{1}{2}g_1(M^2_{\pi^0},T)\right)\right],
\end{equation}
where is worth noting that the neutral pion mass decreases with $T$, contrary to what happens if we do not consider electromagnetic effects, as can be seen if we put $M^2_{\pi^\pm}=M^2_{\pi^0}$ in the last expression.

As for the charged pion we can separate the contributions coming from pion tadpoles, $M^2_{\pi\text{ tadpoles}}$; and the two different contributions from the virtual photon diagrams: one coming from the photon tadpole which is not zero at finite temperature and gives a typical thermal screening contribution, $M^2_{Ph.\text{ tadpole}}$; and the other due to the one-photon exchange, $M^2_{\text{Ph.Exchange}}$.

For the first to ones we get $M^2_{\pi\,\text{ Tadpoles}}=\frac{\hat M^2_{\pi^0}}{2F^2}-4Ze^2g_1(M^2_{\pi^\pm},T)$ and $M^2_{\text{Ph. Tadpole}}=\frac{1}{3}e^2T^2.$
The latter has the typical form of a Debye or screening mass of the electric field in a thermal bath \cite{Kraemmer}, which always grows with $T$.

The one-photon exchange diagram is more complex and its contributions to the real part depend, in general, on the three-momentum of the external pion, which is a direct consequence of the Lorentz symmetry breaking in the thermal bath. The Matsubara sums can be performed in the standard way, before performing the analytic continuation to external continuous frequencies
\begin{eqnarray*}
Re\left(\Sigma_{\text{Ph. Exchange}}\right)=e^2\left(\int\!\!\frac{d^3k}{(2\pi)^3}\frac{n(\omega)}{2\omega}\frac{(2q-k)^2|_{k_0=\omega}}{(q_0-\omega)^2-\omega'^2}
+\int\!\! \frac{d^3k}{(2\pi)^3}\frac{n(\omega)}{2\omega}\frac{(2q-k)^2|_{k_0=-\omega}}{(q_0+\omega)^2-\omega'^2}\right.\\+\left.
\int\!\! \frac{d^3k}{(2\pi)^3}\frac{n(\omega')}{2\omega'}\frac{(2q-k)^2|_{k_0=q_0-\omega'}}{(q_0-\omega')^2-\omega^2}+
\int\!\! \frac{d^3k}{(2\pi)^3}\frac{n(\omega')}{2\omega'}\frac{(2q-k)^2|_{k_0=q_0+\omega'}}{(q_0+\omega')^2-\omega^2}\right),
\end{eqnarray*}
with  $\omega^2=\vec k^2$ the photon energy squared inside the loop and $\omega'^2=(\vec q-\vec k)^2+M^2_{\pi^\pm}$ the pion energy squared, also inside the loop; being $k$ the virtual photon four-momentum. The above three-momentum integrals can be written as
\begin{eqnarray}
Re\left(\Sigma_{\text{Ph. Exchange}}\right)=\frac{e^2}{2\pi^2}\int_0^\infty\!\! dk\int_0^\pi\!\! d\phi\, n(k)\sin\phi\frac{M^2_{\pi^\pm}\sqrt{E^2-M^2_{\pi^\pm}}\cos\phi-k(E^2\sin^2\phi+M^2_{\pi^\pm}\cos^2\phi)}{E^2\sin^2\phi+M^2_{\pi^\pm}\cos^2\phi}\\ \nonumber+
\frac{e^2}{4\pi^2}\int_0^\infty\!\! dk\int_0^\pi\!\! d\phi\, \frac{n(\omega')}{\omega'}\sin\phi\frac{k^2(M^2_{\pi^\pm}\cos^2\phi+E^2\sin^2\phi)+2M^2_{\pi^\pm}(E^2-k\sqrt{E^2-M^2_{\pi^\pm}}\cos\phi)}{E^2\sin^2\phi+M^2_{\pi^\pm}\cos^2\phi},
\end{eqnarray}
It is clear now that the charged pion real part of the self-energy depends on the energy of the external pion $E^2=\vert\vec{q}\vert^2+M_{\pi^\pm}^2$, and, therefore, on the external three-momentum.

\begin{figure}[h!t]
\includegraphics[scale=0.8]{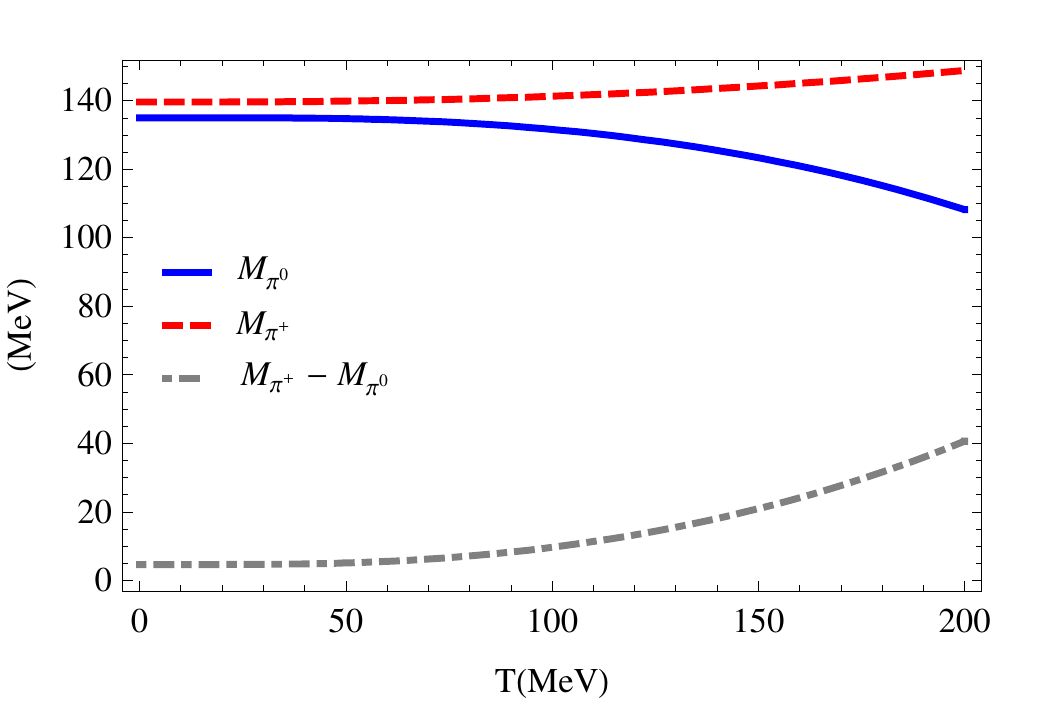} 
\centering
\caption{Preliminary results for the charged (dashed line), neutral (solid line) masses and its difference (dot-dashed line) at LO in the static limit.}
\label{fig:masasdif}
\end{figure}

The final result will take the form
$M^2_{\pi^\pm}=\hat M^2_{\pi^\pm}(T=0)+M^2_{\pi\text{ Tadpoles}}+M^2_{\text{Ph. Tadpole}}+Re\left(\Sigma_{\text{Ph. Exchange}}\right).$ In Figure \ref{fig:masasdif} we have plotted our preliminary results for the neutral and charged pion real masses as a function of the temperature of the thermal bath, taking physical values for the pion masses and calculating $M^2_{\text{Ph.Exchange}}$ in the static limit, i.e, for $E=M_{\pi^\pm}$.
Once we have the thermal and isospin breaking corrections to the masses separately for any value of the external momentum, we can calculate it in the limit  where temperatures are (i) much greater than the masses and the external momenta (which means that we have to set the masses inside the loops to zero), and (ii) sizeable to the momenta running inside the loops, $T\sim k\gg m,q$ . With this assumptions, we are lead to the HTL result given in \cite{Manuel:1998sy},
$M^2_{\pi^\pm}-M^2_{\pi^0}=\hat M^2_{\pi^\pm}\left(1-\frac{T^2}{6}\right)+\frac{1}{4}e^2T^2,$
which serves us as a consistency check.

\begin{figure}[h!t]
\includegraphics[scale=0.8]{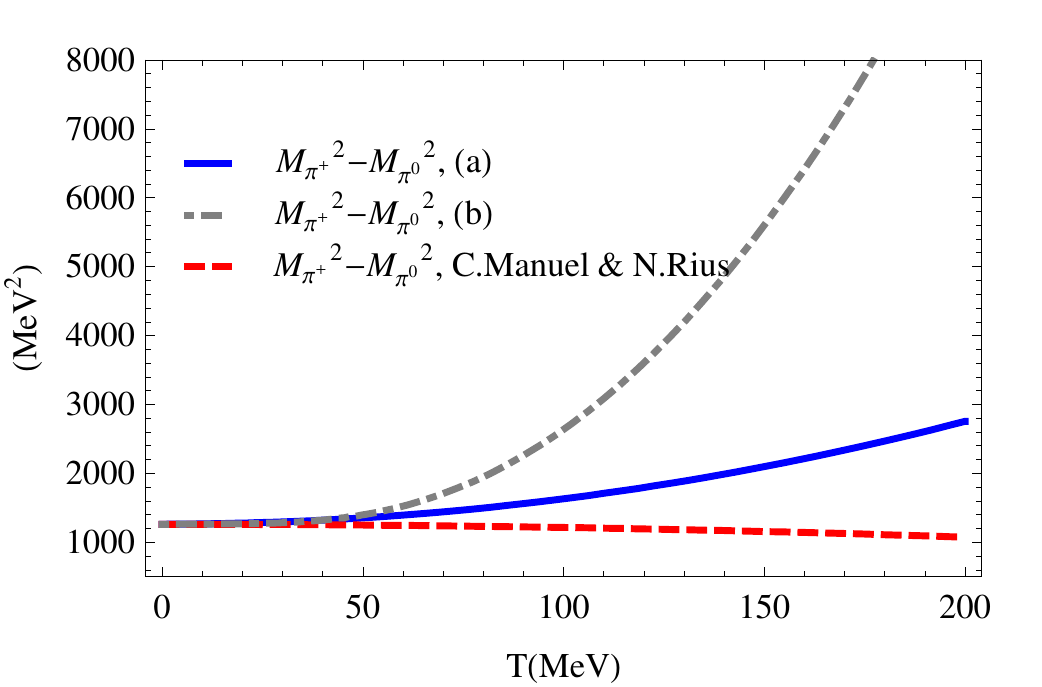} 
\centering
\caption{Different results for the charged-neutral pion mass difference. (a, solid line) corresponds to our preliminary results in the chiral limit keeping corrections $e\neq0$ for the tree level charged pion mass inside the loops, (b, dot-dashed line) corresponds to the same preliminary calculation with $m_u=m_d\neq0$ and $e\neq 0$ also inside the loops; and the full dashed line is the result given in  \cite{Manuel:1998sy}. }
\label{fig:comparacion}
\end{figure}

Moreover, our low temperature analysis allows to assume a slightly different chiral limit, in the sense that we can still assume $m_u=m_d=0$ but considering $e\neq0$ even inside the loops. In Figure \ref{fig:comparacion} we have plotted our preliminary calculation both in this latter limit, and also  considering $m_u=m_d\neq 0$, $e\neq0$; to be compared with those appearing in the HTL result \cite{Manuel:1998sy} where the contributions from the screening-like term, always increasing with $T$ and inherent to the thermal bath, are responsible for the final growth of the mass difference, contrary to what one would expect naively from the sum rule relating the axial and vector spectral functions \cite{Kraemmer} applying chiral symmetry restoration arguments. 

\end{document}